\documentclass{revtex4}
\usepackage{amsmath}
\usepackage{amssymb}
\usepackage{graphicx}

\begin{document}

\title{Decelaration/acceleration phases with the Higgs field}

\author{Vladimir Dzhunushaliev}
\email{vdzhunus@krsu.edu.kg}
\affiliation{Institut f\"ur  Physik, Universit\"at Oldenburg, Postfach 2503, 
D-26111 Oldenburg, Germany; 
\\
Department of Physics and Microelectronic
Engineering, Kyrgyz-Russian Slavic University, Bishkek, Kievskaya Str.
44, 720021, Kyrgyz Republic \\
and \\
Institute of Physics of National Academy of Science
Kyrgyz Republic, 265 a, Chui Street, Bishkek, 720071,  Kyrgyz Republic}

\author{Kairat Myrzakulov}
\affiliation{Dept. Gen. Theor. Phys., Eurasian National University, Astana, 010008, Kazakhstan}

\author{Ratbay Myrzakulov}
\affiliation{Dept. Gen. Theor. Phys., Eurasian National University, Astana, 010008, Kazakhstan
}
\email{cnlpmyra1954@yahoo.com}


\begin{abstract}
It is shown that the Einstein gravity + Higgs scalar field have cosmological regular solutions with deceleration/acceleration phases and with bouncing off from a singularity. The behavior of the solution near to a flex point is in detail considered. 
\end{abstract}

\maketitle

\section{Introduction}

Supernova observations \cite{SN1,SN2} were the first to suggest that our universe is currently accelerating. For this acceleration now it is believed that as much as 2/3 of the total density of the universe is in a form which has large negative pressure and which is usually referred to as dark energy. It has been proposed number of various models aiming at the description of dark energy universe (for review, see \cite{padmanabhan}). 
\par
It is evidently that to have deceleration (where $\ddot a < 0$) and acceleration (where 
$\ddot a > 0$) phases it is necessary to have the moment with $\ddot a = 0$. In this paper we show that the gravitating Higgs scalar field may have cosmological solutions with such property. Such solution exists only with a single value of cosmological constant. 

\section{Numerical solution with deceleration/acceleration phases}

The aim of this section is to show that in the ordinary Einstein gravity interacting with the Higgs scalar field there exist solutions having the deceleration and acceleration phases. The corresponding Einstein + scalar field equations are 
\begin{eqnarray}
	R_{\mu \nu} - \frac{1}{2} g_{\mu \nu} R &=& \varkappa T_{\mu \nu}, 
\label{2-10}\\
	\frac{1}{\sqrt{-g}} \frac{\partial }{\partial x^\mu} 
	\left(
		\sqrt{-g} g^{\mu \nu} \frac{\partial \phi}{\partial x^\nu}
	\right) &=& - \frac{d V(\phi)}{d \phi}
\label{2-20}
\end{eqnarray}
where $V(\phi)$ is 
\begin{equation}
	V(\phi) = \frac{\lambda}{4} \left(
		\phi^2 - m^2
	\right)^2 - V_0 
\label{2-30}
\end{equation}
$\phi$ is the Higgs field and the Lagrangian for the scalar field is 
\begin{eqnarray}
	L &=& \frac{1}{2} \left( \nabla_\mu \phi \right) 
	\left( \nabla^\mu \phi \right) - V(\phi),
\label{2-40}\\
  T_{\mu \nu} &=& \left( \nabla_\mu \phi \right) 
	\left( \nabla_\nu \phi \right) - g_{\mu \nu} L.
\label{2-50}
\end{eqnarray}
The quantity $V_0$ in Eq. \eqref{2-30} is identical to a cosmological constant. Later we will see that $V_0$ is defined uniquely in a flex point $t_0$ where $\ddot{a}(t_0) = 0$. 
\par
\begin{figure}[h]
\begin{minipage}[t]{.45\linewidth}
 \begin{center}
 \fbox{
  	\includegraphics[height=.8\linewidth,width=.8\linewidth]{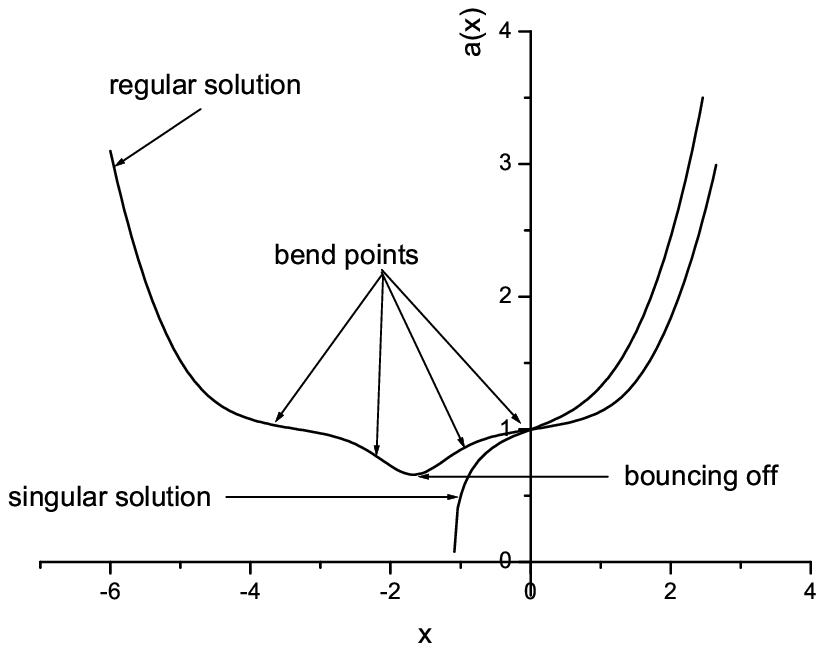}}
  \caption{The profiles $a(t)$ for regular ($\dot a_0 = 0.08$) and singular 
  ($\dot a_0 = 0.2$) solutions. The parameters 
  $\lambda = 0.41, a_0 = 1.0, \phi_0 = -0.0028; 
  \dot \phi_0 = - \frac{\sqrt{2\dot a_0^2 + 2}}{a_0}, m = 0.603$ 			are the same for both solutions.}
  \label{fg1}   
	\end{center}
\end{minipage}\hfill
\begin{minipage}[t]{.45\linewidth}
 \begin{center}
 \fbox{
	  \includegraphics[height=.8\linewidth,width=.8\linewidth]{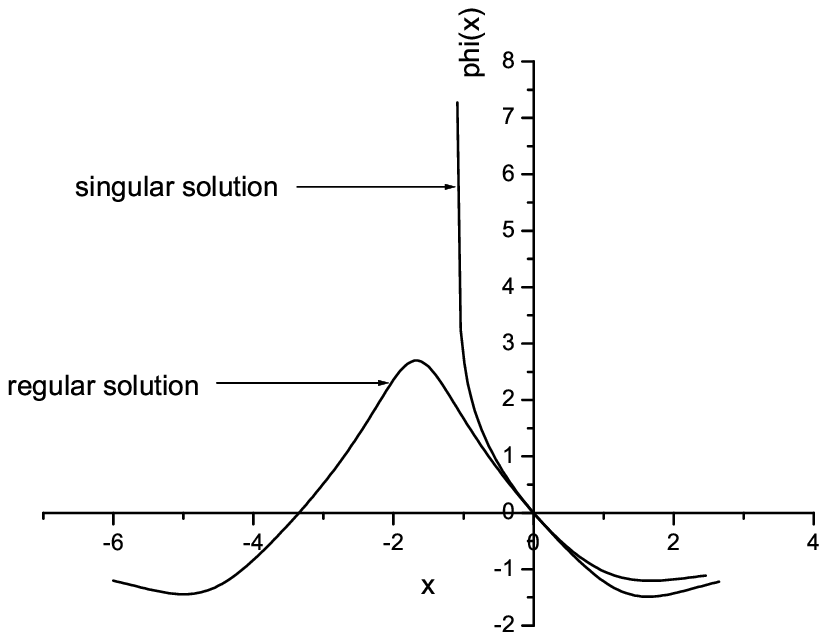}  
 }
	\caption{The profiles $\phi(t)$ for regular and singular solutions.}
	\label{fg2}   
	\end{center}
\end{minipage}\hfill 
\end{figure}

 
We consider the cosmological metric 
\begin{equation}
	ds^2 = dt^2 - a^2(t) \left[
		d \chi^2 + \sin^2 \chi \left(
			d \theta ^2 + \sin^2 \theta d \varphi^2
		\right)
	\right] .
\label{2-60}
\end{equation}
The equations for $a(t)$ and $\phi(t)$ are 
\begin{eqnarray}
	\frac{3 \dot a^2}{a^2} + \frac{3}{a^2} &=& \varkappa 
	\left[ 
		\frac{\dot \phi^2}{2} + \frac{\lambda}{4} 
		\left( \phi^2 - m^2 \right)^2 - V_0	
	\right] ,
\label{2-70}\\
  \frac{2 \ddot a}{a} + \frac{\dot a^2}{a^2} + \frac{1}{a^2} &=& \varkappa  
	\left[ 
		- \frac{\dot \phi^2}{2} + \frac{\lambda}{4} 
		\left( \phi^2 - m^2 \right)^2 - V_0	
	\right] ,
\label{2-80}\\
  \ddot \phi + \frac{3 \dot a \dot \phi}{a} &=& \lambda \phi 
  \left( m^2 - \phi^2 \right).
\label{2-90}
\end{eqnarray}
For the deceleration phase we have $\ddot a(t) < 0$, for the acceleration phase: $\ddot a(t) > 0$. Consequently there is a flex point $t_0$ where 
\begin{equation}
	\ddot a(t_0) = 0. 
\label{2-100}
\end{equation}
Using the condition \eqref{2-100} and equations \eqref{2-70} \eqref{2-80} one can find the following constraints on the initial conditions and cosmological constant 
\begin{eqnarray}
	\frac{2 \dot a^2_0}{a^2_0} + \frac{2}{a^2_0} &=& \varkappa  
	\left[ 
		\frac{\lambda}{4} \left( \phi^2_0 - m^2 \right)^2 - V_0	
	\right] ,
\label{2-110}\\
  \frac{2 \dot a^2_0}{a^2_0} + \frac{2}{a^2_0} &=& \varkappa  
	\dot \phi^2_0
\label{2-120}
\end{eqnarray}
where $a_0 = a(t_0), \dot a_0 = \dot a(t_0), \phi_0 = \phi(0), \dot \phi_0 = \dot \phi(0)$. It means that the cosmological constant is defined uniquely 
\begin{equation}
	\Lambda = \varkappa V_0 =  \frac{\lambda}{4} \varkappa \left( \phi^2_0 - m^2 \right)^2 - 
	2 \left(
		\frac{\dot a_0^2}{a_0^2} + \frac{1}{a_0^2}.
	\right)
\label{2-130}
\end{equation}
For the numerical solution we introduce the dimensionless quantities 
$x = t/\sqrt{\varkappa}$, $\phi \sqrt{\varkappa} \rightarrow \phi$, 
$a/\sqrt{\varkappa} \rightarrow a$. Then the Eq's \eqref{2-70}-\eqref{2-90} are 
\begin{eqnarray}
	\frac{3 \dot a^2}{a^2} + \frac{3}{a^2} &=& \left[ 
		\frac{\dot \phi^2}{2} + \frac{\lambda}{4} 
		\left( \phi^2 - m^2 \right)^2 - V_0	
	\right] ,
\label{2-140}\\
  \frac{2 \ddot a}{a} + \frac{\dot a^2}{a^2} + \frac{1}{a^2} &=& \left[ 
		- \frac{\dot \phi^2}{2} + \frac{\lambda}{4} 
		\left( \phi^2 - m^2 \right)^2 - V_0	
	\right] ,
\label{2-150}\\
  \ddot \phi + \frac{3 \dot a \dot \phi}{a} &=& \lambda \phi 
  \left( m^2 - \phi^2 \right)
\label{2-160}
\end{eqnarray}
with the following initial conditions 
\begin{equation}
	a(0) = a_0, \quad \dot a(0) = a_0, \quad 
	\phi(0) = \phi_0, \quad 
	\dot \phi_0 = - \frac{\sqrt{2 \dot a_0^2 + 2}}{a_0}.
\label{2-170}
\end{equation}
The numerical solution is presented in Fig's \ref{fg1}-\ref{fg2}. We see that thete are two type of solutions: regular and singular one. The regular solution exists for 
$-\infty < t < + \infty$ and has bouncing off point and four flex points. The singular solution has one flex point only. The presented solution is symmetrical one relative to the bouncing off moment but there exist non-symmetrical solutions with the initial conditions different from \eqref{2-170}. 
\par
The asymptotical behavior of the solution is 
\begin{eqnarray}
	a(t) &\approx& a_\infty e^{t \sqrt{\frac{- \varkappa V_0}{3}}} , 
	\quad t \rightarrow + \infty , 
\label{2-180}\\
  a(t) &\approx& a_\infty e^{-t \sqrt{\frac{- \varkappa V_0}{3}}} , 
  \quad t \rightarrow - \infty ,
\label{2-190}\\
  \phi(t) &\approx& -m + \phi_\infty e^{- \alpha t}, \quad 
  \alpha_{1,2} = \sqrt{\frac{-3 \varkappa V_0}{4}} \pm 
  \sqrt{\frac{-3 \varkappa V_0}{4} - 2 \lambda m^2} , 
  \quad |t| \rightarrow \infty 
\label{2-200}
\end{eqnarray}
where $a_\infty, \phi_\infty$ are constants. This solution can describe the inflation of Universe with the posterior standard decay of the scalar field. 

\begin{figure}[h]
 \begin{center}
 \fbox{
	  \includegraphics[height=6cm,width=7cm]{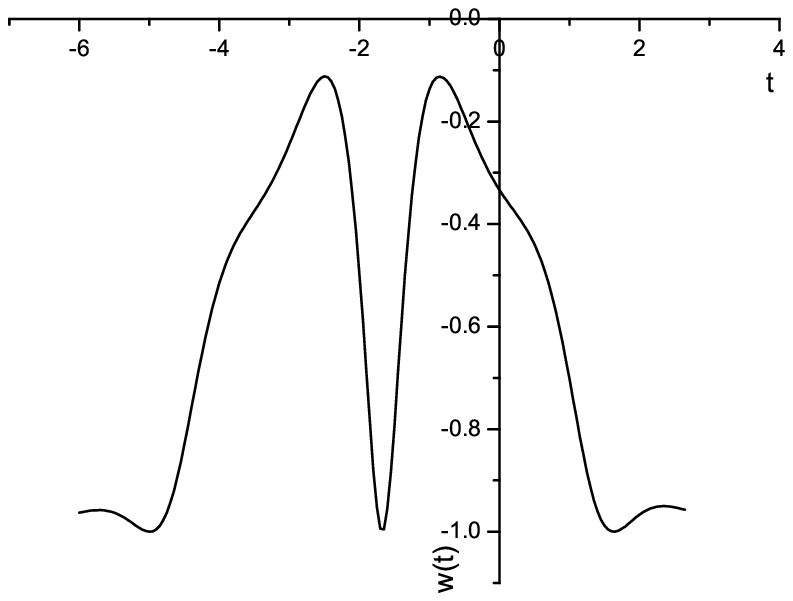}  
 }
	\caption{The profiles $w(t)$ for the regular solution.}
	\label{fg3}   
	\end{center}
\end{figure}

It is useful to present the profile of a state equation (see, Fig. \ref{fg3}) in the form 
\begin{equation}
	w = \frac{p}{\varepsilon} = \frac{T^{11}}{T^{00}} = 
	\frac{\frac{\dot \phi^2}{2} - \frac{\lambda}{4} \left( \phi^2-m^2 \right) + V_0}
	{\frac{\dot \phi^2}{2} + \frac{\lambda}{4} \left( \phi^2-m^2 \right) - V_0}
\label{2-210}
\end{equation}
where $p$ is the pressure and $\varepsilon$ is the energy density. Especially interesting is the behavior of $w$ in the region $x > x_0$, i.e. in the acceleration region. We see that there 
$0 < w < -1$. 

\section{The deceleration $\rightarrow$ acceleration transition}

In the preceding section we have shown that the Universe filled with the Higgs field may have the deceleration/acceleration phases and have presented the solution with 
$a_0 \approx l_{Pl} \propto \sqrt{\varkappa}$. In this section we would like to investigate more carefully the solution near the flex point where the transition from the deceleration to acceleration phase is happened and with 
$a_0 \gg l_{Pl} \propto \sqrt{\varkappa}$. It is not too hard to find the solution of equations set \eqref{2-70}-\eqref{2-90} in the following form 
\begin{eqnarray}
	a(t) &=& a_0 + \dot a_0 t + 
	\left[
		4 \frac{\dot a_0 \left( \dot a_0^2 + 1 \right)}{a_0^2} + 
		\lambda \phi_0 \sqrt{2 \varkappa \left( \dot a_0^2 + 1 \right)}
		\left( m^2 - \phi_0^2 \right)
	\right] \frac{t^3}{6} + \cdots ,
\label{3-10}\\
  \phi(t) &=& \phi_0 - \frac{1}{a_0} \sqrt{2 \frac{\dot a_0^2 + 1}{\varkappa}} t + 
  \left[
  	3 \frac{\dot a_0}{a_0^2} \sqrt{2 \frac{\dot a_0^2 + 1}{\varkappa}} + 
  	\lambda \phi_0 \left( m^2 - \phi_0^2 \right) 
  \right] \frac{t^2}{2} - 
\nonumber \\
	&&
	\left[
		12 \frac{\dot a_0^2}{a_0^3} \sqrt{2 \frac{\dot a_0^2 + 1}{\varkappa}} + 
		3 \lambda \frac{\dot a_0 \phi_0}{a_0} \left( m^2 - \phi_0^2 \right) + 
		\frac{\lambda}{a_0} \sqrt{2 \frac{\dot a_0^2 + 1}{\varkappa}} 
		\left( m^2 - 3 \phi_0^2 \right)
	\right] \frac{t^3}{6} + \cdots 
\label{3-20}
\end{eqnarray}
The deceleration parameter is 
\begin{eqnarray}
	q(t) &=& - \frac{a \ddot a}{\dot a^2} \approx - \frac{a_0 a_3}{\dot a_0^2} t , 
\label{3-30}\\
	a_3 &=& 4 \frac{\dot a_0 \left( \dot a_0^2 + 1 \right)}{a_0^2} + 
					\lambda \phi_0 \sqrt{2 \varkappa \left( \dot a_0^2 + 1 \right)}
					\left( m^2 - \phi_0^2 \right).
\label{3-40}
\end{eqnarray}
More convenient in this approach is the modified deceleration parameter 
\begin{equation}
	q(t) = - \frac{\left( a - a_0 \right) \ddot a}{\dot a^2} \approx 
	- \frac{a_3}{\dot a_0} t^2
\label{3-45}
\end{equation}
The Habble constant is 
\begin{equation}
	H(t) = \frac{\dot a}{a}  \approx \frac{\dot a_0}{a_0} - 
	\frac{\dot a_0^2}{a_0^2} t + \left(
		\frac{\dot a_0^3}{a_0^3} + \frac{a_3}{2}
	\right) t^2 .
\label{3-50}
\end{equation}
Let us remind that the time $t$ is counted from the flex point moment $t_0$. Unfortunately it is not for a while yet unknown: is the solution with $a_0 \gg l_{Pl}$ regular or singular, i.e has the solution bouncing off from a cosmological singularity or no. 

\section{Outlook}

We have shown that the Einstein-Higgs gravity has cosmological solutions with the deceleration/acceleration phases. Additionally these solutions may have bouncing off from a cosmological singularity. The detailed investigation is made near the moment where the transition from the deceleration epoch to the acceleration one happens. 
\par 
One question remains not clear in the carried out research: is the solution with big $a_0$  regular or it has singularity either in the past (Big Bang) or and in the future (Big Rip) or both ?

\section{Acknowledgments} 
I am grateful to the Research Group Linkage Program of the Alexander von Humboldt Foundation for financial support, to J. Kunz for invitation to Universit\"at Oldenburg for research.

\end{document}